# From band reconstruction to Bogoliubov dispersion: How $d_{z^2}$-band enhances iron-based superconductivity

**Jingming Yan[1,2], Shendong Su[1,2], Guihao Jia[1,2], Yucong Peng[1,2], Xuanyu Long[3], Zheng Liu[4], Pei Ouyang[1,2], Qi-Kun Xue[1,2,5,6,7*] and Wei Li[1,2,7*]**

[1] State Key Laboratory of Low-Dimensional Quantum Physics, Department of Physics, Tsinghua University, Beijing 100084, China
[2] Frontier Science Center for Quantum Information, Beijing 100084, China
[3] Department of Materials Science and Engineering, University of Utah, Salt Lake City, UT 84112, USA
[4] School of Physics, Beihang University, Beijing 100191, China
[5] Beijing Academy of Quantum Information Sciences, Beijing 100193, China
[6] Southern University of Science and Technology, Shenzhen 518055, China
[7] Hefei National Laboratory, Hefei 230088, China
[8] Jingming Yan, Shendong Su and Guihao Jia contributed equally to this work.

*Author to whom any correspondence should be addressed.

**E-mail:** qkxue@mail.tsinghua.edu.cn; weili83@tsinghua.edu.cn

## Abstract

Conventional understanding of iron-based superconductivity is usually established based on electronic bands at the Fermi level, leaving deeper bands overlooked as potential factor. Here, we expand this view by demonstrating that a deep-lying $d_{z^2}$ band below the Fermi level can be employed to enhance the superconducting paring strength. By using scanning tunneling microscope to mechanically change the crystal lattice of an iron-based superconductor, we observe an upward shift of the deep-lying $d_{z^2}$ band related feature, leading to hybridization with the primary superconducting bands at the Fermi level. We visualize the novel evidence of this band hybridization and resulting enhanced superconducting gap through quasiparticle interference imaging, providing conclusive evidence that engineering such correlations between deep and pairing-active bands is a potent mechanism for amplifying the superconducting gap. This work refines the conventional view of pairing-relevant electronic structures and establishes band hybridization as an effective way to optimize superconducting materials.

## 1. Introduction

In iron-based superconductors, *d*-orbital bands are crucial for electron pairing [1-19]. The pairing-relevant bands, identified by both experiment and theory [20-22], are $d_{xz}/d_{yz}$ and $d_{xy}$ orbitals (figure 1(a) and (b)), which are located around the Fermi level ($E_F$). The $d_{z^2}$ orbital band, by contrast, resides several hundred meV below $E_F$ (figure 1(b)) and is presumed irrelevant to superconductivity.

A recent study from our group showed that a repulsive STM tip force induces tensile strain in monolayer FeSe/$SrTiO_3$ (FeSe/STO), shifting the $d_{z^2}$ band upward toward the Fermi level ($E_F$) owing to its high sensitivity to the in-plane lattice constant [23-26], and driving its hybridization with the pairing-active $d_{xy}$ band [26] (figure 1(c) and (d)). This process gives rise to a two-stage enhancement of the superconducting gap, increasing from 17.8 meV to 23.6 meV: before the $d_{z^2}$ and $d_{xy}$ bands cross, enhanced electronic correlations boost the gap; upon further tip approach, a stronger enhancement occurs once the two bands intersect.

However, the band evolution depicted in figure 1(d) is a schematic inference from that prior work; direct experimental evidence for the proposed hybridization and its causal link to gap amplification remains lacking. Moreover, the evidence in that work is based solely on local tunneling spectroscopy (STS), which establishes a correlation between gap amplification and band hybridization. Crucially, the spectroscopy technique can only access the local electronic structure, lacking the momentum-space resolution required to reveal the underlying mechanism. Two fundamental questions therefore remain unresolved: whether the enlarged gap is truly superconducting, and whether the proposed band hybridization can be supported by momentum space evidence. Answering these questions is essential to validate band manipulation as a general strategy for tailoring superconducting pairing in the materials with multiple orbital degrees of freedom.

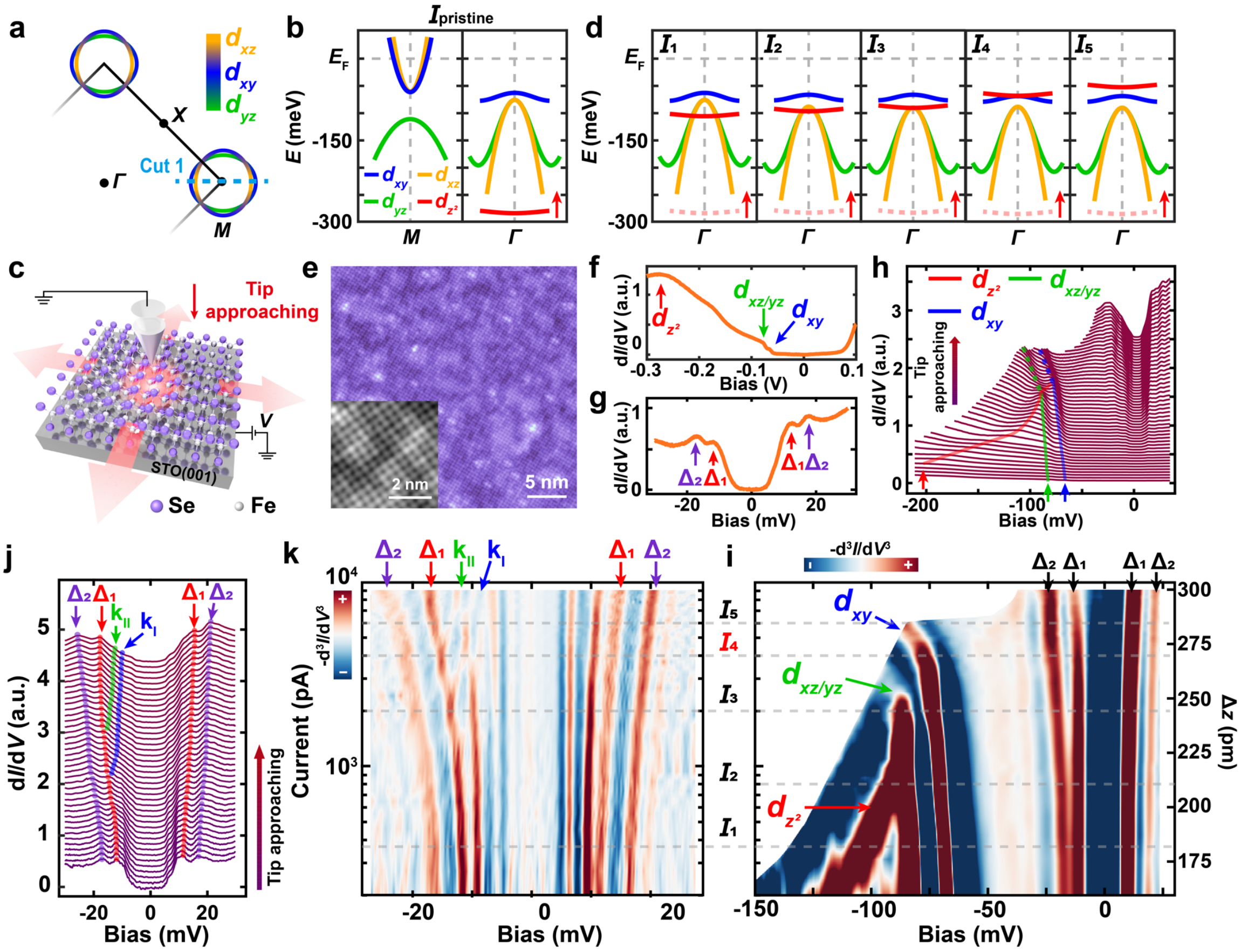

**Figure 1. Tip-induced *d*-orbital band reconstruction and its influence on superconducting coherence peaks in monolayer FeSe/STO.** (a), (b) Schematic of Fermi surface and band structure of monolayer FeSe/STO. Color code indicates orbital characters of bands. Cut 1 denote Γ-M direction. (c) Schematic of experimental setup. Tip-sample spacing is precisely reduced by increasing tunneling current $I_t$. (d) Positions of $d_{z^2}$ band under five tunnelling currents $I_1$-$I_5$. (e) Atomically resolved STM topography of Se-layer (30 nm × 30 nm; set point, $V$ = 30 mV, $I_t$ = 100 pA). Inset: zoom-in topography of Se-lattice (30 mV, 1 nA). (f) A wide-bias range d$I$/d$V$ spectrum (set point, $V$ = 30 mV, $I_t$ = 5 pA). Colored arrows indicate $d_{z^2}$, $d_{xz/yz}$ and $d_{xy}$ bands. (g) d$I$/d$V$ spectrum (30 mV, 500 pA) of a superconducting gap with two pronounced coherence peaks. (h) Current-dependent wide-bias range d$I$/d$V$ spectra (30 mV, $I_t$ from 30 pA to 9 nA), revealing the simultaneous evolutions of *d*-orbital bands and superconducting gaps. Color-matched dashed lines are guides to band shifts. (i)Second derivative [-d²(d$I$/d$V$)/d$V^2$] of (h) ($I_t$ from 100 pA to 9 nA), the corresponding Δ$z$ values are displayed on the right axis (Δ$z$ = 0 is defined at $V$ = 30 mV, $I_t$ = 2 pA). Red-colored regions highlight evolution trajectories of energy bands (arrows) and superconducting gaps (near $E_F$). Gray dashed lines mark five characteristic currents: $I_1$ (0.35 nA), $I_2$ (0.8 nA), $I_3$ (2 nA), $I_4$ (4 nA) and $I_5$ (6 nA), with $I_3$ and $I_4$ denoting the currents for merging of $d_{z^2}$-$d_{xz/yz}$ and $d_{z^2}$-$d_{xy}$ bands. (j) Current-dependent d$I$/d$V$ spectra of superconducting gaps at 0.4 K (30 mV, $I_t$ from 100 pA to 9 nA), showing an increase in coherence peak energies with newly emerged kinks at $I_3$ and $I_4$. Red and purple dots highlight inner and outer superconducting gaps ($\Delta_1$ and $\Delta_2$). Blue and green dots denote kinks emerging from inner coherence peak. (k) Second derivative [-d²(d$I$/d$V$)/d$V^2$] of (j). The kinks are more pronounced.

Here, we directly address both questions by using momentum-resolved quasiparticle interference (QPI) and Bogoliubov QPI (BQPI) measurements. These techniques are built on low-temperature STM platform but provide momentum-space information that local spectroscopy cannot access. Our measurements achieve three advances. First, we visualize the evidence of $d_{z^2}$-orbital-related band hybridization (figure 1(d)) in *k*-space via collective QPI. Second, we map the corresponding superconducting Bogoliubov dispersions and show their characteristic suppression by temperature and magnetic field, thereby establishing the superconducting nature of the amplified gaps. Third, through quantitative analysis of the Bogoliubov QPI patterns, we extract the anisotropic gap function and its evolution under controlled band tuning. Together, these results provide definitive causal evidence that engineering a deep-lying band can decisively enhance the superconducting pairing strength, moving beyond a correlative observation to establish a mechanistic paradigm. In the broader context of lattice-tunable electronic structures and superconductivity in iron-based superconductors [24,27,28], our finding necessitates a revision of the pairing framework and introduces "deep-band engineering" as a general design principle for multiorbital superconductors.

## 2. Results

### 2.1. Spectroscopic evidence for band hybridization and gap enhancement

Figure 1(c) illustrates the experimental configuration for the STM measurement, which enables the precise control of tip-sample distance by adjusting the tunneling current ($I_t$) under a constant bias voltage. High quality monolayer FeSe films were grown on $SrTiO_3$(001) substrate. We focus on a specific region of a monolayer FeSe film figure 1(e), containing some dumbbell-shaped Fe-vacancies served as scattering sources for the QPI. At the setpoint with no tip perturbation, wide-bias range $dI/dV$ spectrum (figure 1(f)) reveals well-defined features of distinct *d*-orbital bands consistent with the ARPES measurements (figure S1, also see the schematic of pristine *d*-bands in figure 1(b)) [22, 23, 26, 29], establishing a baseline for tracking subsequent band shifts. The $dI/dV$ spectrum measured on a defect-free region exhibits a superconducting gap with two pronounced coherence peaks (figure 1(g)). In contrast, the $dI/dV$ spectrum taken on the top of the Fe vacancy (figure S2) manifests a pair of bound states at ± 5 mV.

The Fermi surface of monolayer FeSe/STO comprises electron-like pockets at the M point with dominant $d_{xz}$, $d_{yz}$ and $d_{xy}$ orbital character (figure 1(a)) [21]. Increasing the tunneling current applies a repulsive force that expands the FeSe lattice (figure 1(c)), thus shifting the $d_{z^2}$ band related feature toward $E_F$ (figure 1(b) and (d)) [26].

First, following our previous approach [26], we track the evolution of the *d*-orbital bands through current-dependent wide-bias $dI/dV$ spectra (figure 1(h), (i), S3 and S4). As the tip-sample distance decreases (with increased $I_t$), the $d_{xz/yz}$ and $d_{xy}$ band shift slightly away from $E_F$, whereas the $d_{z^2}$ peak moves rapidly toward $E_F$ and sequentially merges with the $d_{xz/yz}$ and $d_{xy}$ bands at tunneling currents $I_3$ and $I_4$, respectively. Using the strain-sensitive $d_{z^2}$ band as an internal spectroscopic marker, we further provide a rough estimate of the local lattice expansion induced by the tip (Supplementary Note I). The *d*-orbital band hybridization is evidenced by the increased bending of the $d_{xz/yz}$ and $d_{xy}$ bands away from $E_F$ at $I_3$ and $I_4$ (see dashed-lines in figure 1(h), (i)). This hybridization of the $d_{z^2}$ band with the pairing-relevant *d*-orbitals leads to a noticeable enlargement of the superconducting gap near $E_F$ (figure 1(i) and figure S3). Based on the trends in figure 1(i), we select six characteristic tunneling currents ($I_{\text{pristine}}$, $I_1$-$I_5$) to track the $d_{z^2}$ band evolution and illustrate the corresponding schematic band structures at each current (figure 1(b) and (d)). The direct visualization of this band reconstruction at different currents will be presented in the following sections. Note that the schematic in figure 1(d) illustrates only the current-dependent shift of the $d_{z^2}$ band; it does not resolve its precise hybridization with the other $d$-bands, owing to the limited momentum-space resolution of our STM measurements.

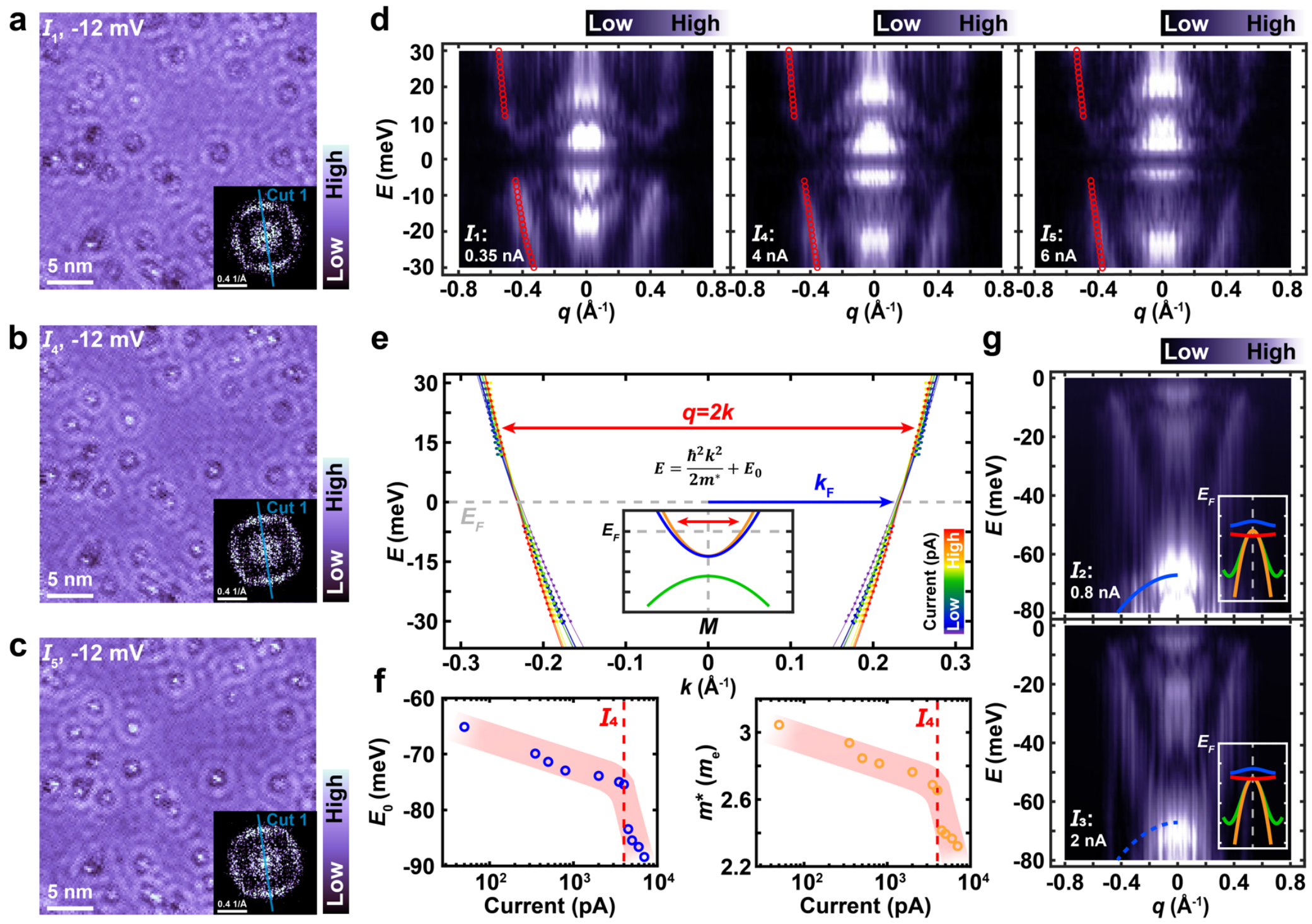

**Figure 2. Band reconstruction under different tunneling currents revealed by quasiparticle interference.** (a)-(c) Real space d$I$/d$V$ maps acquired at -12 mV under different tunneling currents (30 nm × 30 nm; set point, $V$ = 30 mV, $I_1$ = 0.35 nA, $I_4$ = 4 nA, $I_5$ = 6 nA). Insets: Fast Fourier transformation of the corresponding real-space d$I$/d$V$ maps. Cut 1 denote the Γ-M direction (figure 1(a)). (d) Energy-dependent line profiles along Cut 1, from -30 mV to 30 mV. Additional d$I$/d$V$ maps and their corresponding FFT images at different energies are provided in Supplementary Material figure S5-7. Red circles mark the scattering wave vectors obtained from the normal state quasiparticle interference. These wave vectors exhibit an electron-like band dispersion, attributed to intra-band scattering at the M point (see schematic in figure 1b). (e) Fitted electron-like $d$-orbital bands dispersion at the M point under different tunneling currents. The colored circles represent the scattering wave vectors extracted from figure 2(d) and Supplementary Material figure S8. The solid lines are parabolic fitting results and the gray dashed line denotes $E_F$. Inset: schematic of the intra-band scattering near the M point. (f) Current-modulated band bottom energy ($E_0$) and electron effective mass ($m^*$) evolution. Colored guidelines track the evolution of $E_0$ and $m^*$, with their trends exhibiting a transition at $I_4$. (g) Energy-dependent line profiles along Cut 1 at $I_2$ (0.8 nA) and $I_3$ (2 nA) with wide energy range. The blue solid (dashed) line denotes the presence (absence) of the hole-like $d_{xy}$ band. Insets: positions of $d_{z^2}$ band at $I_2$ and $I_3$.

To achieve higher energy resolution, we measured the superconducting gaps at 0.4 K (figure 1(j) and (k)). With increasing tunneling current, the coherence peaks shift to higher energies, signaling a significant gap enhancement. Strikingly, this amplification is accompanied by the emergence of two new distinct kink-features on the coherence peaks (labeled $k_I$ and $k_{II}$ in figure 1(j) and (k)) precisely at currents at $I_3$ and $I_4$, which correspond to the merging of the $d_{z^2}$ band with the $d_{xz/yz}$ and $d_{xy}$ band tops (figure 1(d)), respectively. This clear spectroscopic response provides new evidence that the hybridization of the $d_{z^2}$ band with *each* of the pairing-relevant $d$- bands at higher energies (around –90 meV and –77 meV) directly influences the superconducting Bogoliubov quasiparticles near $E_F$. The origin of the observed kink-features warrants further study; however, the clear increase in the slope of gap growth at both $I_3$ and $I_4$ (figure 1(k)) provides direct spectroscopic evidence that band reconstruction drives the gap amplification.

**2.2. Imaging band reconstruction via QPI**

To visualize the evolution of the electronic bands under tip-induced strain, we perform QPI imaging at different tunneling currents. Real-space d$I$/d$V$ maps acquired within the region marked in figure 1(e) exhibit clear scattering patterns induced by Fe vacancies (figure 2(a)-(c)). Their corresponding fast Fourier transform (FFT) patterns are shown in the insets. The ring-like scattering wave vectors observed in these FFT maps originate from intra-pocket scattering within the electron-like band centered at the M point (figure 1(b)) [28, 29]. Additional d$I$/d$V$ maps and their corresponding FFT images at more energies are provided in figure S5 to S7.

Energy-dependent line profiles along the Γ-M direction are extracted from the FFT data (Cut 1 in figure 2(a)-(c) insets) at selected currents $I_1$, $I_4$ and $I_5$ (figure 2(d)). While the scattering vectors in the normal state maintain a parabolic dispersion, they shift systematically with increasing current. By tracking this evolution across a full set of 11 currents (figure 2(d) and figure S8), we quantitatively extract the intra-pocket scattering wave vectors (Figure 2(e)). Parabolic fitting of the extracted QPI dispersions allows us to determine the band bottom and effective mass of the electron-like pocket at the M-point (figure 2(f) and figure S9). Both parameters exhibit a clear transition near current $I_4$: the band bottom shifts downward while the effective mass decreases, signaling a significant band reconstruction when the $d_{z^2}$ band merges with the band top of $d_{xy}$. Importantly, the observed reduction in effective mass only reflects a local effect, namely, the "pulling down" of the electron-like $d_{xy}$ band near $E_F$ via its hybridization with the upward-shifted $d_{z^2}$ band. Whether this hybridization similarly affects the bandwidth of the $d_{xy}$ and $d_{yz/xz}$ bands across the entire Brillouin zone remains unclear; such a determination would require full momentum-space mapping, which is beyond the scope of the present QPI analysis. The nearly identical normal-state $k_F$ extracted from the fitting curves across all 11 currents (figure 2(e) and Supplementary Material Note II) indicates that the doping concentration remains unchanged. This confirms that band reconstruction is the primary cause of the gap enhancement.

Crucially, the QPI data reveal a second distinct signature at current $I_3$. The hole-like $d_{xy}$ band, visible around -70 meV at lower currents (e.g., $I_2$), disappears once the $d_{z^2}$ band merges with the $d_{xz/yz}$ band top at $I_3$ (figure 2(g)), owing to its intensified hybridization with the continuously upshifting $d_{z^2}$ band. Note that, due to the limited momentum resolution of our QPI measurements, the exact positions of the $d$-orbital bands in the schematic (figure 1(d)) may not be accurate; nevertheless, the evolutionary trends extracted from our data provide a comprehensive picture of the current-dependent band reconstruction. Together, these abrupt changes in the QPI-derived band parameters at $I_3$ and $I_4$ provide momentum-resolved evidence for the band hybridization events previously identified in spectroscopy, linking the $d_{z^2}$- related feature's rise to a reconstruction of the $d$-bands.

**2.3. Bogoliubov dispersion evidence for superconductivity amplification**

To establish the causal link between band reconstruction and the enhanced superconducting state, we turn to Bogoliubov quasiparticle interference (BQPI), a probe of the collective response of superconducting quasiparticles. We perform BQPI measurements at the same characteristic currents ($I_1$, $I_4$ and $I_5$) as in figure 2. The superconducting gap spectra at these currents confirm the progressive amplification of two distinct coherence peaks, corresponding to gap maxima $\Delta_1$ and $\Delta_2$ (figure 3(a)-(c)).

Re-examining the energy-dependent line profiles from the Fourier-transformed QPI patterns along Cut 1 (figure 3(d)-(f)), we observe, in addition to the normal-state intra-pocket scattering, distinct new sets of scattering signals (marked by green and cyan circles) within the gap energy. These are identified as BQPI wave vectors.

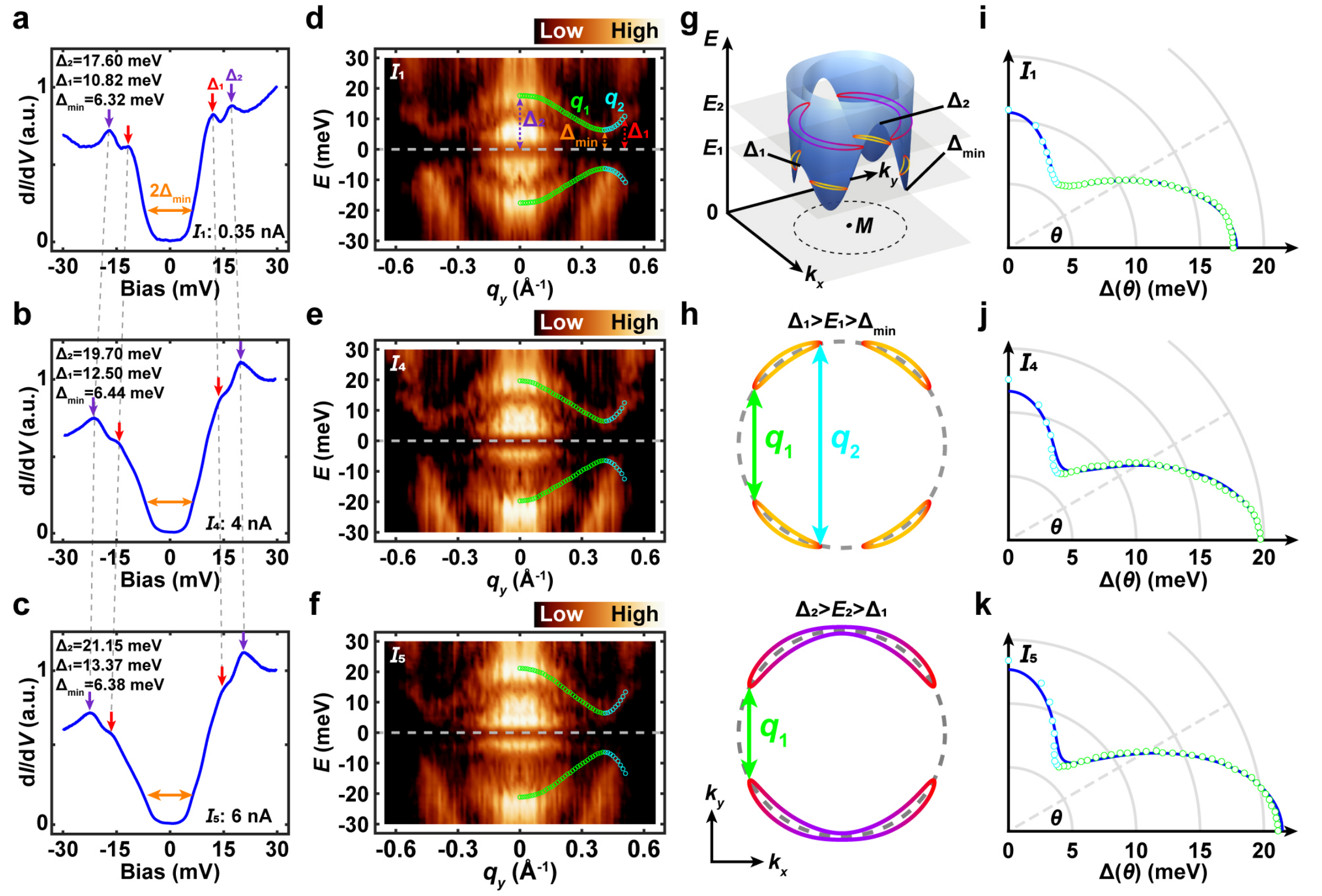

**Figure 3. Bogoliubov quasiparticle interference in monolayer FeSe/STO under different tunneling currents.** (a)-(c) Typical $dI/dV$ spectra of the superconducting gap obtained under the same tunneling currents as in figure 2(a)-(c) (set point, $V$ = 30 mV, $I_1$ = 0.35 nA, $I_4$ = 4 nA, $I_5$ = 6 nA). Red, purple and orange arrows highlight the inner superconducting gap ($\Delta_1$), outer superconducting gap ($\Delta_2$) and gap minimum ($\Delta_{min}$), respectively. Dashed gray lines track the evolution of superconducting gap $\Delta_1$ and $\Delta_2$. (d)-(f) Energy-dependent line profiles along Cut 1 in the insets of figure 2(a)-(c), showing clear Bogoliubov quasiparticle dispersions under different tunneling currents. The BQPI wave vectors $q_1$ and $q_2$ are marked as green and cyan circles. Colored arrows indicate $\Delta_1$, $\Delta_2$ and $\Delta_{min}$. The gray dashed lines denote $E_F$. (g) Schematic of the band structure in the superconducting state at the M point. Normal-state Fermi surface at the M point is shown as the dashed circle at $E$ = 0 meV plane. (h) CECs of Bogoliubov quasiparticles at two typical energies $E_1$ and $E_2$ ($\Delta_{min} < E_1 < \Delta_1$, $\Delta_1 < E_2 < \Delta_2$) extracted from (g). The red color at the tips of the banana-shaped contours represents high spectral weight regions. The scattering wavevector $q_1$ appears within the energy range between $\Delta_2$ and $\Delta_{min}$, while $q_2$ appears within the energy range between $\Delta_1$ and $\Delta_{min}$. (i) to (k) Polar plots of the angular-dependent superconducting gaps extracted from BQPI wave vectors $q_1$ and $q_2$ in (d) to (f), represented by green and cyan data points, respectively. Solid curves are gap fittings using Eq. (1) to the data.

The superconducting gap in monolayer FeSe/STO was found to be nodeless but moderately anisotropic [21, 27]. Its superconducting gap exhibits strong anisotropy in momentum-space, which can be described by the formula [21]:

$$\Delta(\theta) = \Delta_0 + A\cos(2\theta) + B\cos(4\theta) \quad (1)$$

where the polar angle $\theta$ is defined with respect to the center of the M-point electron pocket. The $\cos(2\theta)$ and $\cos(4\theta)$ term represents the angular harmonic component required to capture the twofold and fourfold modulation of the gap anisotropy, respectively. The superconducting gap exhibit two local maximum value $\Delta_1$ and $\Delta_2$ corresponding to the two coherence peaks in figure 3(a)-(c), and a gap minimum $\Delta_{min}$. Such a gap anisotropy can give rise to a typical Bogoliubov band structure (figure 3(g)), manifesting as banana-shaped constant-energy contours (CECs) whose morphology depends on the energy $E$ relative to these gap extrema [16, 32-37]. The derivation of such Bogoliubov band dispersion and CECs is described in Supplementary Note III.

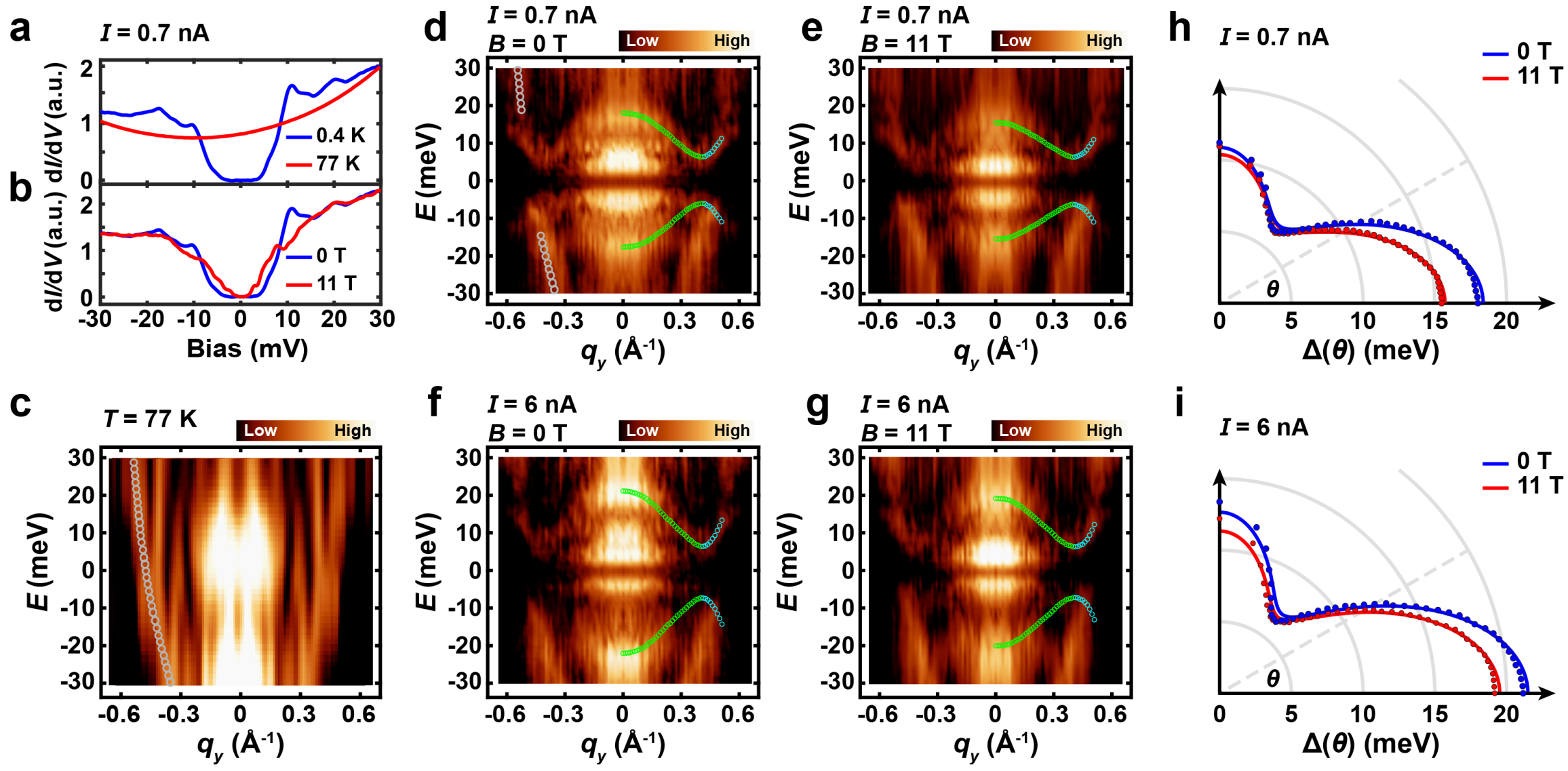

**Figure 4. Temperature and magnetic field response of Bogoliubov quasiparticle interference in monolayer FeSe/STO.** (a) d$I$/d$V$ spectra at 0.7 nA under different temperatures. Spectra in (a) were measured with the setpoint of 0.7 nA ($V$ = 30 mV), under different temperature of 0.4 K and 77 K, respectively. (b) d$I$/d$V$ spectra at 0.7 nA under B = 0 T and 11 T. Spectra in (b) were measured with the setpoint of 0.7 nA ($V$ = 30 mV), under perpendicular magnetic fields of 0 T and 11 T, respectively. (c) Energy-dependent line profiles along Cut 1 at 77 K (set point, $V$ = 30 mV, $I$ = 2 nA). Gray circles mark the scattering wave vectors obtained from the normal state quasiparticle interference. (d)-(g), Energy-dependent line profiles along Cut 1, showing Bogoliubov quasiparticle dispersions as a function of magnetic field and current. (d), (e) were acquired at B = 0 T, for currents of 0.7 nA and 6 nA, respectively. (f), (g) were acquired at B = 11 T, for the same respective currents. The BQPI wave vectors $q_1$ and $q_2$ are marked as green and cyan circles. (h)-(i), Polar plots of the angular-dependent superconducting gaps extracted from BQPI wave vectors $q_1$ and $q_2$ in (d)-(g), represented by colored dots. Solid curves are gap fittings using Eq. (1) to the data in corresponding colors.

We divide these CECs into two classes: $\Delta_{min} < E_1 < \Delta_1$ and $\Delta_1 < E_2 < \Delta_2$, exhibiting four and two energy contours, respectively (figure 3(j)). For a given energy $E$, each contour has two endpoints that are anchored to the underlying Fermi surface where $\Delta = E$. Similar to the octet model in cuprates [32-35], the BQPI wave vectors here connect the endpoints of the CECs which contain high spectral weight (e.g., $q_1$ and $q_2$ in figure 3(h)). For different energies $E_1$ and $E_2$, there exist 10 and 3 possible BQPI wave vectors, respectively (figure S11). Among these vectors, $q_1$ and $q_2$ along the Γ-M direction are crucial for investigating the gap anisotropy (figure 3(h)), as the angular-dependent gap $\Delta(\theta)$ can be sketched from the dispersion of $q_1$ and $q_2$ (Supplementary Note III).

These characteristic BQPI wave vectors are clearly resolved in (figure 3(d)-(f)). As energy increases from $\Delta_{min}$, $q_1$ progressively decreases and terminates at $\Delta_2$, while $q_2$ increases and terminates at $\Delta_1$, a dispersion consistent with the schematic Bogoliubov bands shown in figure 3(g), (h). This result provides a direct visualization of the superconducting Bogoliubov bands, fulfilling a key goal set forth in the introduction.

Critically, a quantitative analysis of the BQPI dispersions allows us to reconstruct the momentum-dependent superconducting gap function Δ(θ) (see Supplementary Note III). The extracted gap profiles at all three currents are well described by the anisotropic form of Eq. (1) (figure 3(i)-(k)), in full agreement with prior ARPES measurements [21]. Most importantly, the gap amplitudes $\Delta_1$ and $\Delta_2$ extracted from this momentum-resolved, collective quasiparticle response agree quantitatively with those measured locally by STS (figure 3(a)-(c)). This concordance provides unambiguous proof that the tip-amplified gap is indeed superconducting in nature. Furthermore, the strain-driven enhancement of both $\Delta_1$ and $\Delta_2$ is independently and consistently quantified by both local STS (figure 3(a)-(c)) and momentum-resolved BQPI (figure 3(i)-(k)), solidifying the direct correspondence between $d_{z^2}$-band-driven reconstruction and the amplification of superconducting order parameter.

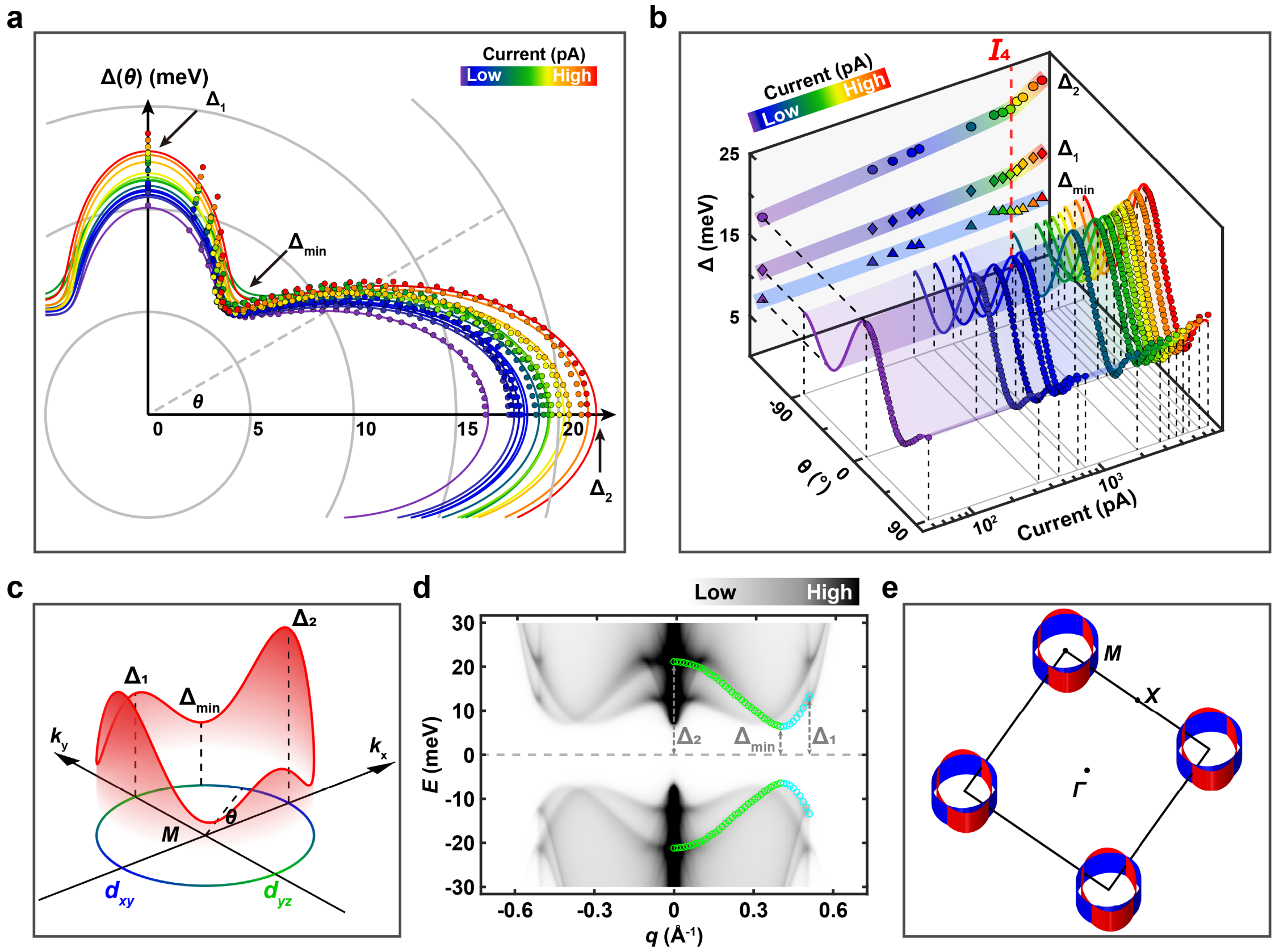

**Figure 5. Tip-tunable superconducting gap anisotropy in monolayer FeSe/STO.** (a) Polar plots of the angular-dependent superconducting gaps extracted from BQPI wave vectors $q_1$ and $q_2$ in figure 3(d)-(f) and Supplementary Material figure S12 under different tunneling currents, represented by colored dots. Solid curves are gap fittings using Eq. (1). $\Delta_1$ and $\Delta_2$ correspond to two local gap maxima, while $\Delta_{min}$ is the gap minimum. (b) The three-dimensional plot of current-dependent gap magnitudes. Two local gap maxima ($\Delta_1$ and $\Delta_2$) and the gap minimum ($\Delta_{min}$) of the fitted curves are projected onto the Δ-current plane. The colored guidelines track the evolution of $\Delta_1$ and $\Delta_2$, exhibiting a clear transition at $I_4$. In contrast, $\Delta_{min}$ remains approximately constant with increasing tunneling current. (c) Schematic of the gap distribution on the electron pockets at the M point. (d) Joint-density-of-states (JDOS) simulation of the scattering intensity with wave vector $q$ parallel to Γ-M direction. Experimentally extracted scattering wave vectors $q_1$ and $q_2$ from figure 3(f) are depicted as green and cyan circles for comparison. (e) Folded Fermi surface of monolayer FeSe/STO.

### 2.4. Temperature and magnetic field responses

To further confirm the superconducting origin of the observed Bogoliubov dispersions, we examine their response to temperature and magnetic field. The local superconducting gap is fully suppressed at 77 K (figure 4(a)) and substantially reduced under an 11 T perpendicular magnetic field (figure 4(b)). Correspondingly, momentum-resolved Bogoliubov quasiparticle dispersions vanish completely at 77 K (figure 4(c)). Under 11 T, these dispersions are strongly suppressed for both weakly (figure 4(d) and (e), $I_t$= 0.7 nA) and strongly (figure 4(f), (g), $I_t$=6 nA) enhanced superconducting states. The angular-dependent gap $\Delta(\theta)$ extracted from the BQPI wave vectors clearly shows this suppression (figure 4(h), (i)). This consistent suppression with both elevated temperature and magnetic field provides definitive evidence that the visualized dispersions originate from superconducting Bogoliubov quasiparticles.

**2.5. Gap evolution and anisotropy from BQPI**

To comprehensively characterize the enlarged superconducting gaps, we extract the angular-dependent gap function directly from BQPI wave vectors $q_1$ and $q_2$ across 13 distinct currents (figure 3(d)-(f) and figure S12). All datasets are well described by Eq. (1), indicating that the gap anisotropy remains nearly unchanged during amplification (figure 5(a)). As the tip approaches, both local gap maxima $\Delta_1$ and $\Delta_2$ increase significantly, with a pronounced upturn in slope occurring at $I_4$ (figure 5(b)), marking a clear transition point.

Mapping the superconducting gap onto the elliptical electron pocket at the M point (figure 5(c)) enables a joint-density-of-states (JDOS) simulation of scattering intensity for wave vectors along Γ-M (figure 5(d) and Supplementary Note IV). The resulting intensity maxima reproduce the Bogoliubov dispersion obtained from our BQPI measurements (green and cyan circles in figure 5(d)).

Notably, while $\Delta_1$ and $\Delta_2$ increase substantially, the global gap minimum $\Delta_{min}$ remains nearly constant across all currents (figure 5(a), (b)). Several mechanisms may contribute to this behavior: (1) orbital-selective Cooper pairing [16, 29, 38], in which coherence is suppressed for certain *d*-orbital bands due to modified spin susceptibility; (2) weak inter-pocket hybridization between the two sign-changing electron pockets at the M point (figure 5(e)), which can produce gap minima [20, 21, 39]; or (3) competition between intra-orbital and inter-orbital pairing [40]. The origin of the persistent $\Delta_{min}$ under gap amplification warrants further investigation.

## 3. Disscussion

From the perspective of collective Bogoliubov quasiparticles, the evolution of the superconducting state under tip-induced strain reveals a two-stage gap enhancement mechanism, demarcated at the current $I_4$ where the $d_{z^2}$ and $d_{xy}$ bands merge (figure 5(b)). The initial enhancement (below $I_4$) is likely driven by increased electronic correlations through tensile strain [26, 41-43], as well as by the renormalization of the primary pairing-active *d*-bands. Below $I_4$, a distinct kink ($k_I$) is observed in the local STS coherence peaks at $I_3$ (figure 1(j), (k)), corresponding to the merger of $d_{z^2}$ band and the top of $d_{xz/yz}$ band (figure 2(d)). However, the slope of the gap increase does not change abruptly in figure 5(b), which may be attributed to the more limited sensitivity of BQPI compared with local STS. In contrast, the second and more pronounced stage (above $I_4$) is directly triggered by the $d_{z^2}$-$d_{xy}$ band hybridization, evidenced both in local STS (kink $k_{II}$, figure 1(k)) and in BQPI (figure 5(b)), underscoring the pivotal role of this previously overlooked orbital channel. Strain-dependent calculations further support this picture, showing that the ARPES-renormalized $d_{z^2}$ band shifts markedly around both Γ and M and hybridizes with the pairing-active bands near M, thereby reconstructing its dispersion (Supplementary Material figure. S13–S15). We note that these calculations are only intended to provide a qualitative description of the strain-induced band evolution, as an accurate theoretical description of the full electronic structure of monolayer FeSe/STO remains challenging. Furthurmore, possible alternative origins of the observed spectral evolution are assessed and ruled out in Supplementary Note V.

The correlation between the *d*-orbital band evolution and superconducting-gap enhancement is further supported by spatially resolved analysis. Using a 10 × 10 grid of tunneling spectra over a 20 nm$^2$ area (figure S18(a)), we extracted both the local $d_{z^2}$-band energy and the outer superconducting gap $\Delta_2$ from the same spatial position under different tunneling currents (figure S18(b,c)). At low tunneling current (0.1 nA), the $d_{z^2}$-band energy exhibits a broad spatial distribution with a variation of approximately 30 meV across the measured area (figure S19), while no corresponding variation in the superconducting gap is observed. At high tunneling current (4 nA), the $d_{z^2}$ band shifts upward, its spatial variation becomes substantially reduced, and the gap still exhibits a variation of approximately 3 meV, indicating that the local gap magnitude is influenced not exclusively by the $d_{z^2}$-band energy but also by other factors such as microscopic inhomogeneity. Although the absolute $d_{z^2}$-band energy does not exhibit a direct spatial correlation with the local gap size at a fixed current, their evolution with increasing current remains systematic: the $d_{z^2}$ band progressively shifts toward $E_F$, while the superconducting gap is overall enhanced (figure S19). Importantly, the current-dependent spectroscopy at different locations further reproduces the same characteristic $d_{z^2}$-band evolution and superconducting-gap enhancement at each position (figure S20).

## 4. Conclusion

Our study integrates momentum-resolved quasiparticle scattering imaging with local spectroscopy to establish a link between $d_{z^2}$-band-driven band reconstruction and the amplification of superconducting gap in monolayer FeSe/STO. Moving beyond the correlative evidence provided by local tunneling spectroscopy, we employ QPI to support in momentum space the tip-induced hybridization between the elevated $d_{z^2}$ band and the pairing-relevant $d_{xy}$ and $d_{xz/yz}$ bands. This constitutes compelling evidence of the proposed band reconstruction. Furthermore, by resolving the collective BQPI, we map the corresponding superconducting Bogoliubov band dispersions. Crucially, the quantitative gap function $\Delta(\theta)$ extracted from these collective excitations matches in both amplitude and anisotropy the data from local STS and prior ARPES measurements [21]. This agreement across local and momentum-resolved probes provides unambiguous evidence that the tip-enhanced spectral gap is indeed a superconducting gap, resolving a key question raised at the outset.

More broadly, our findings demonstrate that engineering correlations between deep, nominally inert electronic bands and the bands at $E_F$ is a potent strategy for enhancing the superconducting pairing strength. This "deep-band engineering" extends the conventional wisdom that only Fermi-surface orbitals matter for pairing. It opens a new avenue for material design and control in iron-based superconductors and potentially other correlated materials where multiple orbital degrees of freedom are present.

## 5. Methods

### 5.1. Sample preparation

High-quality monolayer FeSe films were grown by molecular beam epitaxy method. The Nb-doped (0.05 wt%) $SrTiO_3$(001) substrates (Shinkosha) were degassed in ultra-high vacuum chamber (base pressure is better than $3 \times 10^{-10}$ Torr) at 600 °C for several hours and subsequently annealed at 1150 °C for 20 min to obtain the $TiO_2$-terminated surface. Then, high-purity Fe (99.995%) and Se (99.9999%) sources were co-evaporated by two Knudsen cells to grow monolayer FeSe films on the $SrTiO_3$ substrates. During deposition, the substrates were maintained at 430 °C by applying DC current. The as-grown samples were subjected to stepwise annealing from 450 to 550 °C for hours to optimize sample quality.

### 5.2. STM measurements

Our experiments were carried out on an ultrahigh-vacuum commercial STM system (Unisoku 1300) that can reach a base temperature of 400 mK by using a single-shot $^3$He cryostat. A polycrystalline PtIr STM tip was used and calibrated using Ag island. STS data were taken by standard lock-in method. The feedback loop is disrupted during data acquisition and the frequency of oscillation signal is 973.0 Hz.

### 5.3. DFT calculation

The density functional theory (DFT) calculations are performed with respect to bulk FeSe by using the Vienna *ab initio* Simulation Package. The experimental lattice parameters are used.

The plane-wave cutoff is 500 eV in combination with the projector augmented wave method. The Monkhorst-Pack k-point grid is 12 × 12 × 6. The exchange and correlation is treated by using the Perdew–Burke–Ernzerhof GGA functional. The convergence criteria is $10^{-6}$ eV for electronic iterations.

## Acknowledgements

We thank J. P. Hu, Y. Zhang, J. C. S. Davis and H. Yao for helpful discussions. The experimental work was supported by the National Natural Science Foundation of China (NNSFC, Grant No. 92365201), the National Key R&D Program of China (Grant No. 2022YFA1403103), NNSFC (Grants No. 52388201, and No. 11427903), Quantum Science and Technology-National Science and Technology Major Project (2021ZD0302402) and NSFC (No. 12374062).

## Data availability statement

The data that support the findings of this study are available from the corresponding author upon reasonable request.

## Author contributions

W.L. and Q.-K.X. supervised the project. J.Y., S.S., G.J., Y.P., and P.O. performed the experiments. J.Y. and G.J. grew the samples. W.L., J.Y., S.S. and Y.P. analyzed the data. Z.L. and X.L. performed the DFT calculation. W.L, J.Y. and S.S. wrote the manuscript.

## Conflict of interest

The authors declare no competing financial interests.